# Numerical Simulation of a Quasi-Tropical Cyclone over the Black Sea

V. V. Efimov, M. V. Shokurov, and D. A. Yarovaya
Marine Hydrophysical Institute, National Academy of Sciences of Ukraine, Kapitanskaya ul. 2, Sevastopol, 99011 Ukraine
e-mail: efimov@alpha.mhi.iuf.net

**Abstract**

The paper describes results of numerical experiments on the simulation of a mesoscale quasi-tropical cyclone, a rare event for the Black Sea, with the MM5 regional atmospheric circulation model. General characteristics of the cyclone and its evolution and physical formation mechanisms are discussed. The balances of the momentum components have been estimated, and sensitivity experiments have been performed. It is shown that, according to its main physical properties and energy supply mechanisms, the cyclone can be related to quasi-tropical cyclones.

**Introduction**

The investigation of the formation and development of tropical cyclones and the forecast of the track of their motion are among important areas of research in geophysics. The main background conditions necessary for the formation of tropical cyclones have been studied, a notion about the main physical processes that determine the evolution of tropical cyclones has been formulated, and numerical models have been developed to predict the tracks of cyclones with an increasingly higher lead time and accuracy [1]. However, before the advent of remote sensing methods for the study of the atmosphere and ocean and the onset of satellite cloud imagery, tropical cyclones were regarded as pure tropical natural phenomena.

As is well known, tropical cyclones are classified into a separate group of cyclones that differ from midlatitude lows by their origin, development, and some features of the structure. Tropical cyclones normally have a relatively small size, about 200 to 300 km in diameter, with a central pressure as low as 950 hPa or sometimes even lower than 900 hPa. Wind speeds in spiral bands can reach 70 – 90 m/s. At the center of a tropical cyclone, there is a region 20 – 30 km across with clear or nearly clear skies and weak winds, which is called an eye of the tropical cyclone. The ring surrounding the eye is called the eyewall, and it is here that intense penetrative cumulus convection, heaviest precipitation and thunderstorms, storm winds, and high wind-induced waves occur. Subsidence in the eye produces clear skies there.

Almost all tropical cyclones form in the tropics equatorward of 30° latitudes. With the onset of satellite methods, however, it was found that cyclones similar in structure to tropical ones form from time to time in the extratropical regions, in particular, over the Mediterranean Sea. Favorable conditions for cyclone formation are generally created by outbreaks of North Atlantic air and subsequent cold air-mass transport over the warm sea. This, in turn, produces intense heat fluxes from the sea surface and deep convection.

For example, a small quasi-tropical cyclone, with intense convection and anomalous precipitation, developed from a cold upper-level trough over the western Mediterranean in September 1996 [3]. Two subsynoptic cyclones over the central Mediterranean in early October 1996 also brought heavy precipitation and floods [4]. Quasi-tropical properties of these cyclones — an eye, spiral cloud bands, a warm core, and a typical barotropic structure — were analyzed, and it was shown that the cyclones developed from upper-level cutoff troughs. A quasi-tropical cyclone formed over southeastern Italy in January 1995; an important role in its formation was played by surface heat fluxes [5 – 7]. A mechanism necessary for quasi-tropical cyclones in the Mediterranean to develop from a cold upper-level cutoff cyclone was illustrated by the same example in [8]. A cyclone with hurricane-force winds originated from a filling synoptic depression over southern Italy in January 1982 [9]. Results of the numerical simulation of this cyclone with a detailed description of its structure are given in [10]. In late March 1999, a deep baroclinic cyclone over the Mediterranean generated cumulus convection over vast areas. In the final stage of evolution, it acquired properties of a tropical cyclone: the eye and severe winds in the ring around the eye [11].

Similar mesoscale cyclones, called polar lows, are observed over the ocean at high latitudes. A polar low has small sizes, from several tens to several hundreds of kilometers, and its lifetime does not exceed one and a half days [12]. These intense lows with winds in excess of 15 m/s form over the Atlantic and Pacific oceans between 50° and 70° latitudes in both hemispheres during cold outbreaks from land. Dozens of polar mesocyclones per year form in this region [13]. A characteristic feature of the polar low in satellite images is a twisted spiral region of high convective clouds in the shape of a comma, with a distinct cloud-free eye at the center. Sometimes, polar lows have properties of classical midlatitude cyclones: the initial growth due to baroclinic instability and surface fronts. At the same time, the sensible and latent heat fluxes from the ocean surface and the low-level convergence of moisture fluxes, which causes forced convection and the release of heat of condensation aloft, may play an important role in the formation of polar lows. In this respect, the polar lows are similar to tropical cyclones [14].

An anomalous intense mesoscale cyclone, which resembled a tropical hurricane in appearance, formed over the Black Sea in late September 2005. In this paper, it will be shown that this rare mesoscale cyclone, which developed in the southwestern part of the sea, can be related by its main properties to quasi-tropical cyclones.

Section 1 describes observations of the cyclone, its impact on the sea, and typical features of a synoptic situation. Section 2 provides a brief description of a numerical model used for the simulation of the cyclone. Sections 3 and 4 present the evolution and structure of the cyclone from the numerical simulation. Section 5 shows the estimates of the balances of momentum components. Section 6 contains a description of numerical experiments on the sensitivity of the model to different parametrizations of physical processes. A summary is given in the final section.

**1. Cyclone observation**

A quasi-tropical cyclone over the Black Sea was detected in satellite images over September 25 – 29, 2005. It had a cloud-free eye and distinct spiral cloud bands and was no more than 300 km in diameter. As can be judged from a cloud-top temperature of 223 – 240 K, it was a high cyclone extending to the tropopause. Winds in the zone covered by the cyclone were 20 – 25 m/s according to the QuikScat satellite data. Although the cyclone acquired no devastating characteristics of its tropical counterpart, it induced a sharp deterioration of weather, with delays of cruises from the Crimea and Odessa to Istanbul. The cyclone stayed over the Black Sea from 25 to 29 September slightly wandering, began to move southward on September 29, and entirely left the Black Sea area by September 30.

This atmospheric cyclone had a large influence on the thermal structure of the upper layer of the Black Sea. The cyclonic vorticity of the velocity field of surface wind over the water produced the Ekman divergence in the upper mixed layer of the sea and the rise of the thermocline, even its outcrop to the surface, a decrease in surface temperature, and sea level fall. The QuikScat thermocline rise velocity was $2 \times 10^{-4}$ m/s, while a typical seasonal mean vertical velocity in the upper layer of the Black Sea is $(1 – 2) \times 10^{-6}$ m/s from different estimates. According to satellite data on September 29, the sea surface temperature under the cyclone dropped sharply, by more than 10°C, which suggests the rise of cold waters from a depth of 30 m to the surface. This cold spot persisted for a long time: the temperature contrast was 14°C on September 29, 3 – 4°C on October 13, and 1 – 2°C even on October 23. Altimetric satellite measurements also showed a 25-cm level fall in the region over which the cyclone stayed. Preliminary results of the numerical simulation of the Black Sea circulation for that period showed that the Ekman divergence in the upper mixed layer pushed the warm surface water to a semicircular southwest coast and raised the sea level there, lowering it under the cyclone. The seasonally typical level difference between the shore and the center of the west gyre of 20 cm increased by another 25 cm owing to the action of the cyclone, with the result that the speed of the Black Sea Rim Current in the southwestern part of the sea doubled to a value in excess of 1 m/s [16].

The determination of the causes of the origin and growth of an anomalous cyclone over the Black Sea calls for a detailed study. However, large-scale specific features of the synoptic situation favorable to the origin of the cyclone can be found preliminarily from the NCEP/NCAR operational numerical analysis data [17].

Since September 19, a blocking high persisted over European Russia and stayed there until the end of September. A large-scale cutoff upper-level low with a cold core formed over Spain on September 18. It moved eastward, inducing deep convection over the Mediterranean Sea because of the cold advection onto the warm sea surface. From September 20 to 24, it was blocked over the Balkans by the above-mentioned high. By September 25, this low had filled and a general synoptic situation in the region was characterized by a very weak circulation at all levels in the free atmosphere. Figure 1a shows sea-level pressure and wind fields from the NCEP/NCAR operational analysis with a resolution of 1° × 1° at 00:00 September 25 (hereinafter, all times are GMT). Over the Black Sea, there is a shallow depression with a pressure drop of 4 hPa and a wind speed of 10 m/s. The geopotentials and wind speeds at 850 and 300 hPa are shown in Figs. 1b and 1c for the same time. It can be seen that the blocking high was favorable to weak atmospheric circulation in the Black Sea region.

It is well known that the absence of strong winds and especially of vertical wind shear is a necessary condition for the formation of a tropical cyclone. In the given case, this condition was fulfilled. Because of the absence of strong winds, the cyclone was able to stay long in the same place over the warm sea, and the absence of wind shear favored the development of its barotropic structure.

Another favorable condition for the onset of a tropical cyclone is the low-level background convergence, which collects warm moist air in one place. As can be seen from Fig. 1a, a dipole pattern of surface circulation comprising the anticyclone in the north and a weak cyclone over the sea induced surface convergence over the Black Sea.

An additional favorable factor was a large overheating of the sea surface relative to the surrounding land, which facilitated the increase in moisture content over the sea. The sea – land temperature contrast on September 18 was 7°C (land 17°C, sea 24°C). By September 24, it had reached 12°C (land 11°C, sea 23°C).

The high sea surface temperature and the high air humidity resulted in a decreased atmospheric stability. The vertical profiles of air temperature and dew point at 00:00 September 25 in the western and eastern parts of the sea from the operational analysis data are shown in Figs. 2a and 2b. The thick dashed line shows a change with height in the temperature of an air parcel that rises adiabatically from the sea surface first along the dry adiabat and then, after reaching the condensation level, along the moist adiabat. The solid right-hand curve is the temperature profile, and the one at the left is the dew-point profile. Between the 900-hPa level and virtually up to the tropopause above 300 hPa, the air parcel rising from the surface is warmer than its surroundings. Such atmospheric stratification is favorable to the development of cumulus convection. A measure of instability is the convective available potential energy (CAPE), numerically equal to the area between the dashed and solid curves from the condensation level to the equilibrium level in Fig. 1c [18]. The CAPE distribution for the entire Black Sea region is shown in Fig. 2c. It can be seen that the CAPE reaches large values over the sea, with a maximum of 1600 J/kg.

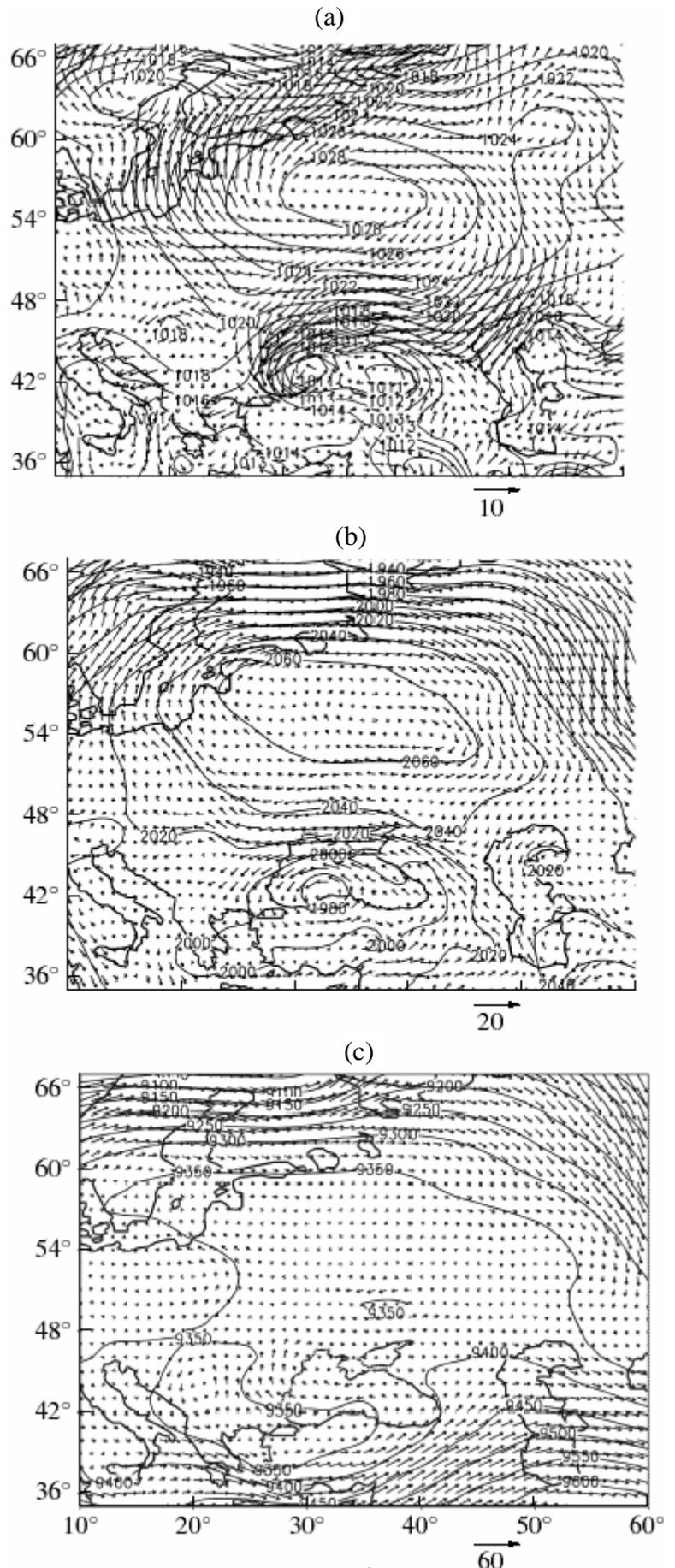

Fig. 1. (a) Wind speed (m/s) at 1000 hPa and sea-level pressure (hPa); (b) geopotential (m) and wind speed (m/s) at 850 hPa; (c) geopotential (m) and wind speed (m/s) at 300 hPa at 00:00 September 25, 2005, from the operational analysis.

It is also seen from Fig. 2a that the dew point is close to the air temperature in the entire troposphere, which is indicative of high relative humidity. This means that the cumulus convection with the release of latent heat of condensation may involve not only the warm moist air from the sea surface, but also the entire moist tropospheric column.

Figure 2b shows sounding profiles for the eastern part of the sea. In contrast to the western part of the sea, in this region, there is a barrier layer up to the 800-hPa level, which prevents convection and in which an ascending air parcel is colder than its surroundings. The measure of intensity of the barrier layer is the negative convective available potential energy (CIN) numerically equal to the area between the solid and dashed curves from the condensation level to the level of free convection in Fig. 2b [18]. The distribution of CIN for the region in Fig. 2d shows that the barrier layer is absent in the southwestern part of the sea. This probably explains why the cyclone developed in that area.

Thus, a vast reservoir of convective available potential energy was located over the entire area of the Black Sea on September 25. The main causes of its formation were the warm surface of the sea and a relatively cold air mass that had formed in the Balkan low during the preceding few days.

**2. Description of the model**

Measurements of mesoscale atmospheric processes are usually scarce. This also applies to the Black Sea cyclone of interest. Apart from QuikScat winds and satellite cloud imagery, there are actually no other high-resolution measurements for this cyclone. A numerical simulation is therefore the main tool to study this mesocyclone.

Version 3.6.2 of the PSU/NCAR mesoscale nonhydrostatic fifth-generation model (MM5) was used for simulation [19, 20]. This model, designed to simulate or predict the mesoscale atmospheric circulation, was adapted to the conditions of the Black Sea region. Skipping the description of MM5, we list only its main capabilities:

(1) multiply nested-grid simulation with a two-way interactive data exchange among the neighboring domains;

(2) nonhydrostatic equations, which allow the model to be used for the simulation of phenomena on a horizontal scale of a few kilometers;

(3) numerous schemes of parametrization of physical processes.

Subgrid-scale processes of horizontal diffusion, vertical momentum, heat, and moisture fluxes, clouds, and precipitation are parametrized in several variants depending on the choice of a spatial resolution.

The model is based on the system of primitive hydrodynamic equations. For integration in time, the leapfrog model with a smoothing Robert – Asselin time filter is used. The assimilation of boundary conditions from the operational analysis outputs into the MM5 model is carried out by a relaxation method: in transition from the boundary of the domain inward, the variable relaxes toward its internal value.

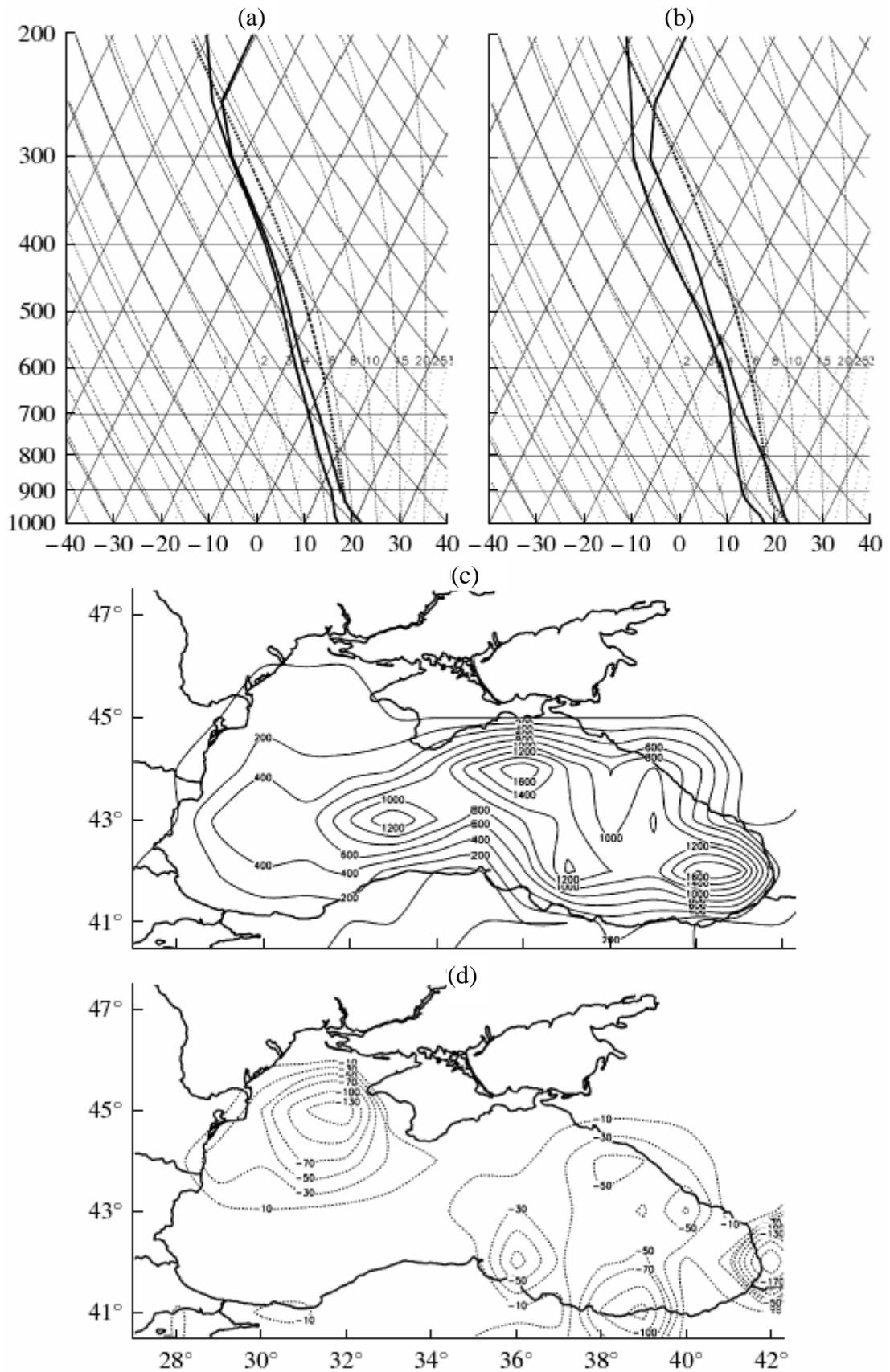

Fig. 2. Vertical soundings at points (a) 32° E, 43° N and (b) 38° E, 43° N and spatial distribution of (c) convective available potential energy (CAPE) (J/kg) and (d) negative convective available potential energy (CIN) (J/kg) from the operational analysis at 00:00 September 25, 2005.

This paper presents results of the simulation on three nested domains with resolutions of 90, 30, and 10 km and sizes of 34 × 31, 64 × 46, and 124 × 73, respectively (Fig. 3). The outermost domain is centered over the Crimean Peninsula, with coordinates of the center at 35° N, 45° E. The size of the domain was chosen sufficiently large to minimize the influence of the boundary conditions on the evolution of the simulated cyclone. In the vertical, the model had 23 irregularly spaced sigma levels, with a higher resolution in the lower troposphere. The initial and boundary conditions were chosen from the NCEP/NCAR global operational analysis, with a spatial resolution of 1° and a time step of 6 h [17]. Thus, the initial conditions for all domains were chosen from the operational analysis, and the boundary conditions were updated every six hours.

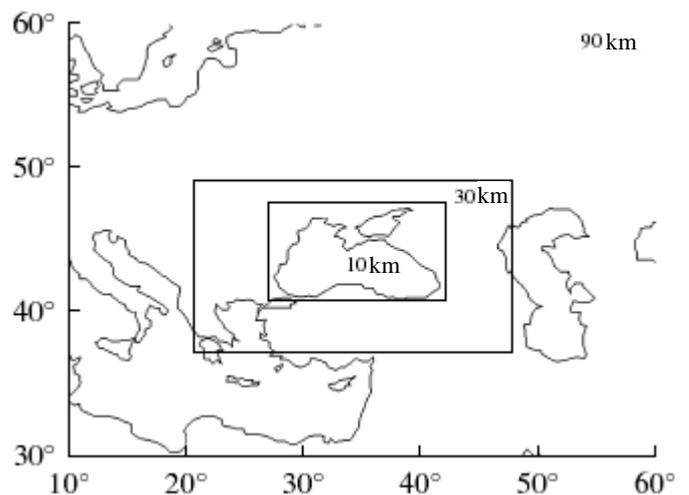

Fig. 3. Location of model domains.
Key: 1. km

The following parametrization schemes were used:

The medium-range forecast (MRF) scheme of parametrization of the planetary boundary layer. The MRF scheme has four stability modes: stable (nighttime) mode, damped dynamic turbulence, forced convection, and free convection. The surface boundary layer is calculated using the Monin – Oboukhov similarity theory. The finite-difference scheme ensures the conservation of mass, energy, and potential entropy.

The rapid radiative transfer model (RRTM) of longwave radiation transfer in the computation of the radiation balance. The parametrization of radiation transfer includes the interaction of short- and longwave radiation with the atmosphere in the presence and absence of clouds.

The simple ice scheme (Dudhia). It calculates microphysical processes of the phase transitions of water in the atmosphere.

The Kain-Fritsch cumulus parametrization scheme for domains with a resolution of 90 and 30 km and the Grell scheme for a domain with a resolution of 10 km. The Grell scheme is best suited for the high-resolution domains.

The land surface temperature was modeled using the equation of heat conduction in the soil, and the sea surface temperature was specified as an external parameter and was not varied during the simulation.

A more detailed description of the parametrization schemes is available in [19, 20].

Initially designed for high-resolution simulations of weather, the MM5 model is now being used for the simulation of tropical cyclones, as well as quasi-tropical Mediterranean cyclones and polar lows [10, 21, 22]. The results of simulations with a resolution of up to 3 km agree well with available data, for example, with radar measurements of precipitation and humidity. The cause of the success of MM5 appears to lie in a correct detailed parametrization of subgrid-scale physical processes.

## 3. Evolution of the cyclone

Because the cyclone to be examined is a rapidly growing unstable disturbance, small variations in the initial conditions may produce large errors in its further evolution. A series of numerical experiments has been conducted, starting at different time from 00:00 September 20 to 00:00 September 25 with a step of 12 h. As might be expected, the modeling result depends heavily on the specification of initial conditions. Without going into details, is may be suggested that the most suitable time to start the simulation of this cyclone is 00:00 September 25, 2005.

An important condition for the generation of a model cyclone appears to be the presence by that time in the operational analysis of a "seed," a weak near-round vortex with a wind speed of about 10 m/s (Fig. 1a), from which a quasi-tropical cyclone could develop afterward. In the simulation of tropical cyclones, a similar "seed" is usually incorporated artificially into the large-scale synoptic environment, because a tropical cyclone is unable to originate without it [22]. In our case, there was no need to do this because the operational analysis model reproduced, though roughly, the "seed" and a mature cyclone developed from it in the high-resolution MM5 model.

As a support of the above, it can be noted that the model failed to reproduce a quasi-tropical cyclone when the initial and boundary conditions were chosen from the NCEP/NCAR analysis with an even coarser 2° resolution. This may be explained by the absence of a "seed" in the initial conditions. Afterward, the variant was considered with the initial conditions at 00:00 September 25 and with the boundary conditions taken from the operational analysis. The model was integrated for six days, from September 25 to September 30.

The model has reproduced well all stages in the life cycle of the cyclone, i.e., the formation stage, the mature stage, and the decay stage after the landfall over Turkey. Figure 4 shows (a) a satellite cloud photograph, (b) 10-m wind speed, (c) sea-level pressure at 12:00 September 27 from the model, and (d) the track of the cyclone center from September 25 to September 29. It is evident from Fig. 4a that the cyclone has an eye and spiral cloud bands, and Figs. 4b and 4c demonstrate that the model-generated cyclone is approximately circular in shape. It can be seen that the model has correctly reproduced the size and position of the cyclone.

Figure 5a displays the time variation of the central pressure $p_{min}(t)$ from September 25 to September 29 and of the maximum 10-m wind speed $V_{max}(t)$, and Fig. 5b shows $R_{max}(t)$, the variation of the radius of maximum winds, i.e., the distance from the cyclone center at which the azimuthally averaged surface wind speed reaches its maximum. Several stages of its development can be identified.

In the initial stage from 00:00 September 25 to 12:00 September 26, the maximum speed of surface wind was about 15 m/s, the central pressure was about 1010 hPa, and the radius $R_{max}$ reached 100 – 115 km. The cyclone had strong asymmetry with distinct spiral irregular bands, in which convection with high vertical velocities and large cyclonic vorticity of the velocity field were concentrated, and was located near the ground. As the cyclone developed, its height was increasing and by the end of the initial stage reached the 700-hPa level. The height of the cyclone was determined by the presence of a well-pronounced azimuthal circulation.

During the early stage, the main source of energy for tropical cyclones is the convective available potential energy acquired from the surrounding water area owing to convergence. The mechanism of intensification in this stage is associated with the release of latent heat of condensation and with surface friction. The release of latent heat of condensation produces the buoyancy force, the air moves upward, the low-level convergence of the velocity field develops, and the vorticity increases because the vortex tubes contract. Due to friction, this cyclonic vorticity causes additional convergence in the boundary layer, an additional increase in the vertical velocity, an increase in vorticity, and so on. Such an intensification mechanism with a positive feedback of convective instability of the second kind (CISK) was proposed to explain the development of an axisymmetric cyclone [23, 24]. At present, the CISK mechanism has been detected in a relatively disorganized array of convective cells developing in the formation stage of tropical cyclones [25]. For the Black Sea cyclone, analysis of satellite images and comparison of the modeling results with tropical cyclone observations suggest that its development during the initial stage was induced by the CISK mechanism.

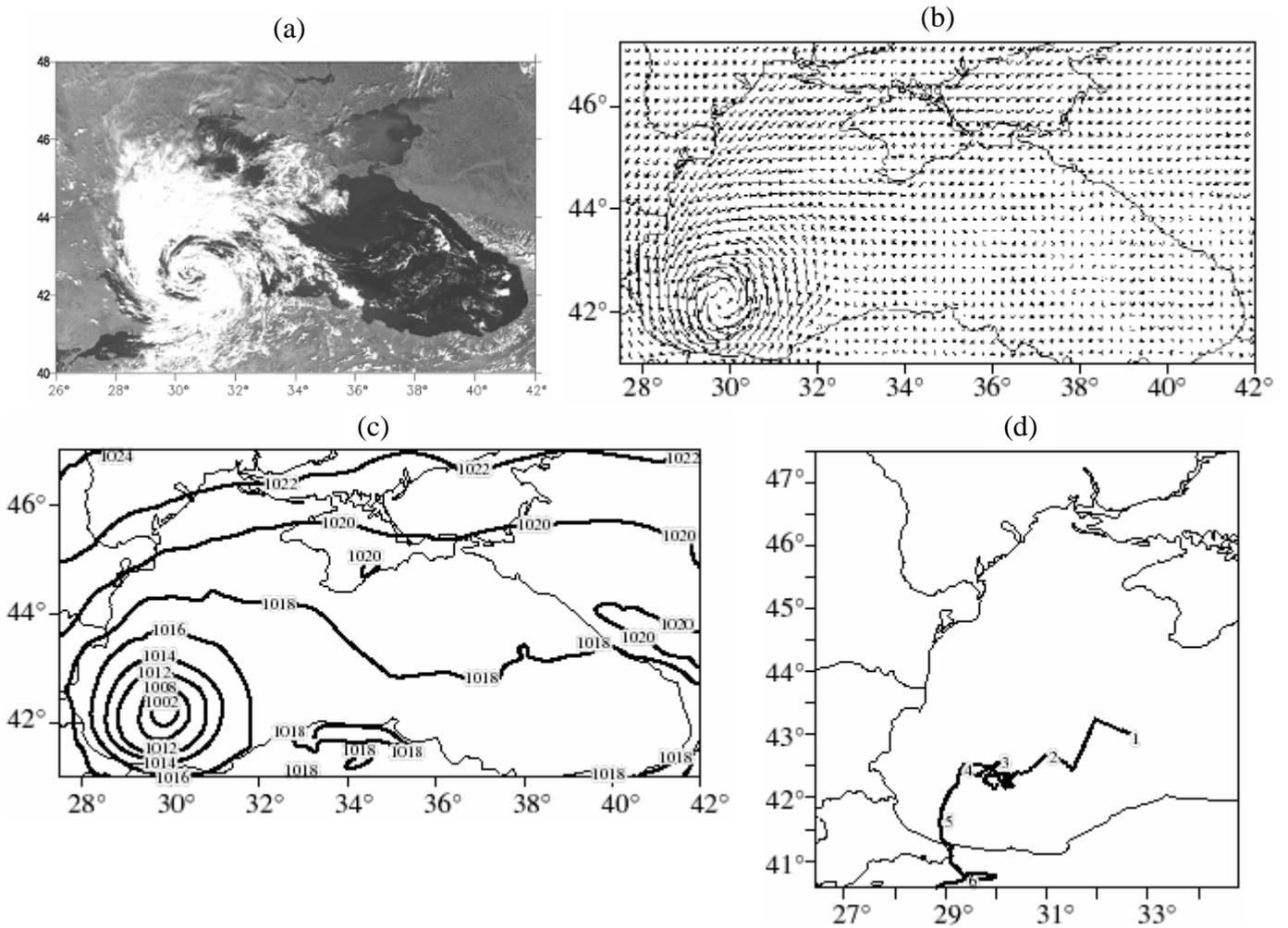

Fig. 4. Comparison of satellite data and the simulation results: (a) satellite-derived clouds and (b) simulated 10-m wind speed (m/s) and (c) sea-level pressure (hPa) at 12:00 September 27. (d) The track of the cyclone center over the Black Sea. The numbers denote the following times: (1) 25.00, (2) 25.12, (3) 26.21, (4) 28.06, (5) 28.21, and (6) 29.12.

The second stage is the rapid development of the cyclone from 12:00 September 26 to 12:00 September 27. Within 24 h, the wind speed increased to 24 m/s, the central pressure fell to 999 hPa, and $R_{max}$ decreased to 65 km. Thus, the cyclone contracted to half of its size and intensified significantly. In addition, the cyclone became more axisymmetric and developed in height substantially, extending to the 300-hPa level.

In a tropical cyclone with a sufficiently strong intensity of the vortex, of importance is a second intensification mechanism, wind-induced surface heat exchange (WISHE) [21, 23, 24]. In this mechanism, as in the CISK mechanism, surface friction leads to convergence; however, the heat source for buoyancy intensification is now not a reservoir of convective available potential energy, but the surface fluxes of sensible and latent heat, which increase substantially with wind speed. The maximum fluxes of sensible and latent heat in the Black Sea cyclone reached values of 300 and 700 W m$^{-2}$ by the end of the second stage.

The third stage is the quasi-stationary developed cyclone from 12:00 September 27 to 12:00 September 28. At this time, the cyclone slightly intensified and central pressure fell to 992 hPa. During this stage, the cyclone was almost circular with a constant radius $R_{max} = 65$ km, and its height did not vary either, remaining at 300 hPa. Eventually, the last stage is the rapid filling and decay of the cyclone from 12:00 September 28 to 00:00 September 29, when it began to approach the shore and made landfall.

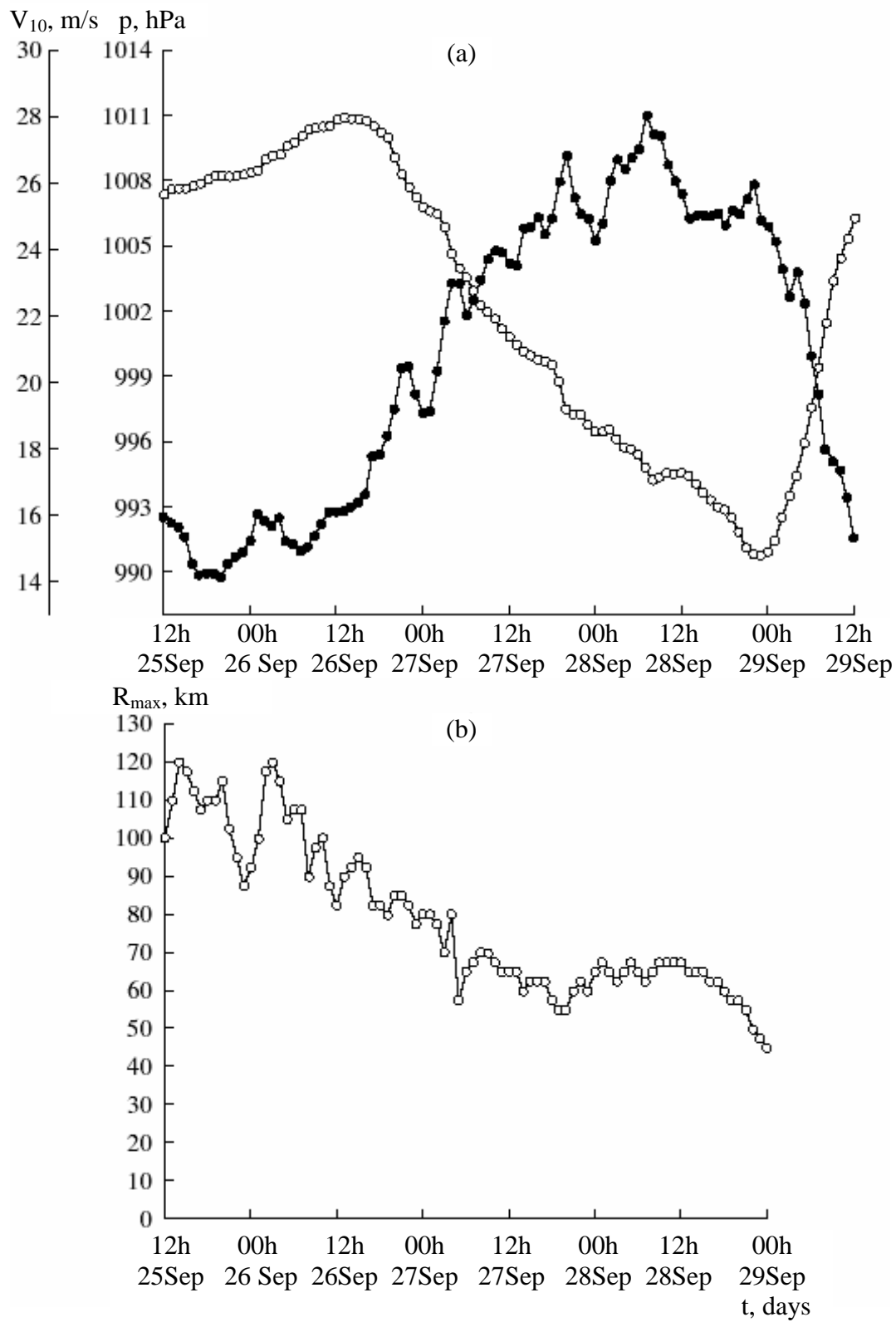

Fig. 5. (a) Temporal variation of central pressure $p_{min}(t)$, hPa (open circles) and maximum 10-m wind speed $V_{max}(t)$, m/s (solid circles); (b) temporal variation of the radius of maximum wind $R_{max}(t)$, km.
Key: 1. m/s; 2. hPa; 3. Sept.; 4. @t, day; 5. km

A detailed study of all the formation and decay stages requires an analysis of sources, sinks, and the rates of change of momentum, angular momentum, vorticity, potential vorticity, kinetic, potential, and thermal energy, and moisture. In this study, we discuss the properties of a steady mature stage when the rates of change of the indicated characteristics are small and balance conditions in the equations for these characteristics are fulfilled, when sources and sinks are balanced. The cyclone achieved the mature stage at 12:00 September 27; by this time, the wind speed reached a maximum of 25 m/s, the radius of the cyclone decreased to a minimum of 65 km, and the cyclone extended to the 300-hPa level. Later on, these parameters remained practically unchanged. The central pressure of the cyclone still continued to fall slightly, but the cause of this fall is not clear (Fig. 5a). The next section shows the vertical structure of the cyclone in the mature stage at 12:00 September 27.

## 4. Structure of the cyclone

*Kinematics*

It is evident from Fig. 4 that the cyclone has an axial symmetry. Therefore, we consider its axisymmetric structure. For this purpose, we use the cylindrical coordinates (r, θ, z) with origin placed at the surface center of a cyclone and moving with it. All variables are averaged over the azimuthal angle θ, so that the resulting azimuthal means depend only on the radius r and height z. The velocity vector in cylindrical coordinates has the azimuthal (tangential) component $V_\theta$, the radial component $V_r$, and the vertical component w. To describe the structure of the cyclone, we introduce the radius of maximum winds $R_{max}$, the distance from the center of the cyclone at which the azimuthally averaged azimuthal wind velocity reaches its maximum; in contrast to the previous section, $R_{max}$ depends now on z. Another important characteristic used for the description of the dynamics of a cyclone is the absolute angular momentum per unit mass $M=V_\theta r+fr^2/2$, the sum of the relative angular momentum $V_\theta r$ and momentum related to the planetary rotation $fr^2/2$, where f is the Coriolis parameter. In the absence of friction, the absolute angular momentum is a Lagrangian invariant, i.e., it is conserved for a moving air parcel.

Eventually, Fig. 6 shows the axisymmetric structure of the cyclone: (a) the azimuthal velocity component $V_\theta$, (b) the radial velocity component $V_r$, (c) the vertical velocity w, and (d) the absolute angular momentum M, all averaged over the azimuthal angle, at 12:00 September 27.

The main feature of a mature hurricane is a ring of extremely strong azimuthal winds. This cyclonic azimuthal circulation is usually called primary. The primary circulation has its maximum at the ground and decreases with height. The azimuthal velocity in tropical cyclones can reach a maximum value of 70 – 90 m/s at $R_{max}$ = 20 – 40 km [1, 2, 22, 27].

The azimuthal velocity distribution $V_\theta(r, z)$ for the Black Sea cyclone is shown in Fig. 6a. At a specified height, $V_\theta$ grows almost linearly as r increases from 0 to $R_{max}$ = 65 km, which corresponds to a solid rotation or constant vorticity, and then decreases with r. As the height increases, $V_\theta$ increases to a maximum value of 27 m/s at 1 km, the 925-hPa level, and decreases above this level. This pattern is qualitatively similar to the distribution for a typical tropical cyclone.

A second important feature of the tropical cyclone is the low-level convergence of the velocity fields, air rise in the eyewall coinciding in position with the region of the maximum azimuthal wind, upper-level divergence, and subsidence on the periphery of the cyclone. This toroidal circulation is usually referred to as the secondary circulation. In an intense tropical cyclone, the radial velocity $V_r$ may reach 25 m/s at lower levels in the inflow region and 12 m/s aloft in the outflow region. The vertical velocity of air rise in the eyewall reaches 2 m/s [1, 2, 22, 27].

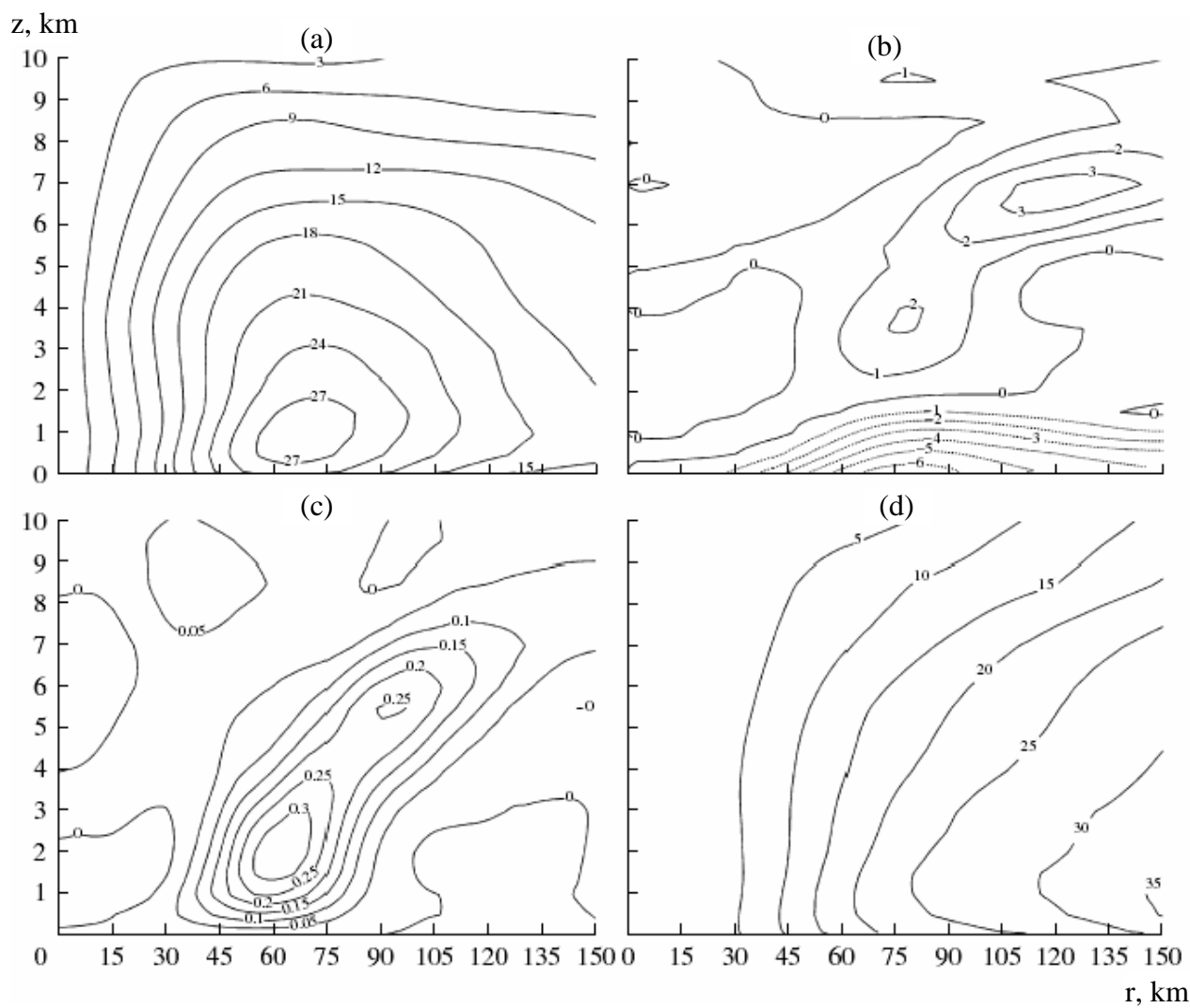

Fig. 6. Vertical cross sections of the azimuthally averaged fields at 12:00 September 27: (a) tangential wind velocity (m/s), (b) radial wind velocity (m/s), (c) vertical wind velocity (m/s), and (d) absolute angular momentum ($10^5$ m$^2$/s). The abscissa is the distance from the cyclone center in km, and the ordinate is the height in km. Solid (dashed) lines correspond to positive (negative) values.
Key: 1. km

The radial and vertical velocity distributions $V_r(r, z)$ and $w(r, z)$ for the Black Sea cyclone are shown in Figs. 6b and 6c. They are also qualitatively similar to the respective distributions for the tropical cyclone. Strong inflow toward the cyclone center, or convergence, occurs in the boundary layer at pressure levels below 2 km, i.e., the 850-hPa level, and strong outflow, or divergence, is observed above 5 km, the 500-hPa level (Fig. 6b). The inflow velocity reaches a maximum value of 5 m/s at a height of 300 m, and the maximum outflow velocity of 3 m/s occurs at 7400 m. In Fig. 6c, one can note the intense vertical rise of air in the eyewall at a distance of 60 km from the cyclone center throughout the troposphere, with a maximum velocity of 0.3 m/s at a height of 2 km at the 700-hPa level, and a weak subsidence in the eye with a maximum velocity of 0.06 m/s.

The main mechanism for producing the primary circulation is the generation of vertical vorticity owing to the stretching of vortex tubes [2, 27]. The low-level vertical air column, with planetary vorticity on the periphery of the cyclone, stretches owing to convergence as it approaches the center of the cyclone, so that the column's vorticity increases. In other words, this means the conservation of absolute angular momentum in the process of low-level convergence.

The ring of the rotating air contracts owing to convergence, and the azimuthal velocity intensifies. If $V_\theta = 0$ at the edge of the cyclone at the distance R from the center, the law of conservation of the absolute angular momentum gives $M = V_\theta r + fr^2/2 = fR^2/2 = $ const. With decreasing r, the azimuthal velocity $V_\theta$ must increase. Moreover, nearer to the center, the ring of air rises upward because of its stretching in the vertical, the convergence and radial velocity decrease rapidly starting at the radius $R_{max}$, and the azimuthal velocity reaches its maximum at r = $R_{max}$. The relationship between $R_{max}$ and $V_{\theta max}$ can be roughly estimated assuming that convergence begins to develop from the shore of the sea, i.e., at R = 200 km; then, $V_{\theta max} = f(R^2 - R_{max}^2)/2R_{max} = 27$ m/s, which coincides with the real value. At upper levels, the ring stretches because of divergence, so that the rotational velocity $V_\theta$ drops and becomes even negative [1, 2]. Indeed, the vorticity in tropical cyclones is negative on the periphery at upper levels.

At lower levels, the absolute angular momentum is not conserved because of friction. As can be seen from Fig. 6d, it drops with a decrease in r. At upper levels, the absolute angular momentum is conserved exactly, and the lines of M follow the streamlines of the secondary circulation more closely.

The mechanisms for producing the secondary circulation are associated with heating, in addition to friction; therefore, we discuss a thermodynamic axisymmetric structure of the cyclone first.

*Thermodynamics*

In a typical tropical cyclone, intense cumulus convection is concentrated in the eyewall. In this region, specific and relative humidity is extremely high and the density of hydrometeors, such as cloud particles, raindrops, cloud ice, snow, and graupel, reaches large values. Precipitation rates also reach maximum values in the eyewall. Outside the eyewall, clouds and precipitation are organized in several spiral bands. The eye itself is often nearly free of clouds, and the air in the eye is very dry owing to subsidence [2, 28].

There was a similar picture in the Black Sea cyclone. Convection was intense in all stages of the cyclone. Convective clouds and convective precipitation were concentrated in the eyewall with a radius of about 60 km. The azimuthally averaged distribution of the hydrometeor density $q_h(r, z)$ and of specific humidity $q(r, z)$ at 12:00 September 27 is shown in Figs. 7a and 7b. For specific humidity, its deviations $\Delta q(r, z)$ from the horizontal mean are also shown. As can be seen, the density of hydrometeors has maximum values of 0.7 g/kg in the slantwise eyewall. The anomaly of specific humidity reaches a maximum in the eyewall, with large values aloft. In contrast, the air in the eye is dry. A similar distribution for relative humidity shows that the relative humidity is above 95% in the eyewall and less than 15% in the upper part of the eye. The horizontal cross sections at different levels show that clouds and precipitation outside the eyewall are organized in spiral bands, with a typical radius of the entire cloud system equal to 150 km, which corresponds to that in Fig. 4a. The rate of precipitation in the eyewall is 10 cm/day in the order of magnitude.

A tropical cyclone, unlike the midlatitude low, normally has a warm core. For vigorous tropical hurricanes, the temperature anomaly at the center of the hurricane relative to its environment reaches 16°C, with a maximum value at a level of 200 to 400 hPa [22]. The Black Sea cyclone also has a warm core. Figure 7c displays the distribution of the azimuthally averaged potential temperature $\theta(r, z)$ and of its deviation from the horizontal mean $\Delta\theta(r, z)$ at 12:00 September 27. As can be seen from Fig. 7c, the cyclone has a warm core with the maximum potential-temperature anomaly $\Delta\theta = 4.5$ K in the eye at a height of 6 km, the 50-hPa level. The temperature anomaly in the warm core is $\Delta T = 3$°C. The warm core is the cause of the lower sea-level pressure at the center of the cyclone relative to its periphery. To a first approximation, this pressure anomaly may be estimated from hydrostatics, which is fulfilled in the Black Sea cyclone as will be shown below.

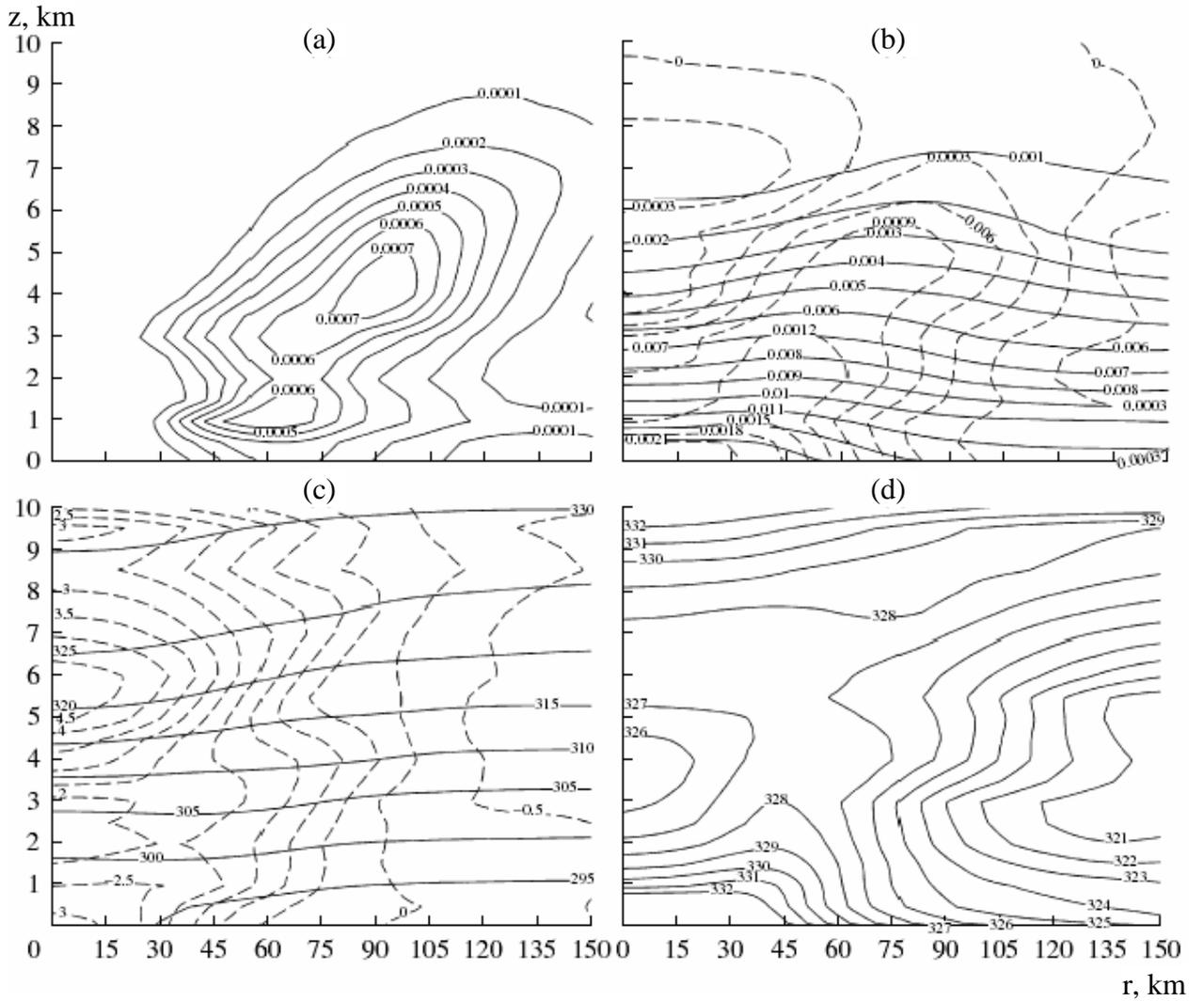

Fig. 7. Vertical cross sections of azimuthally averaged fields at 12:00 September 27: (a) hydrometeor density (kg/kg), (b) specific humidity (solid) and anomaly of specific humidity (dashed) (kg/kg), (c) potential temperature (solid) and anomaly of potential temperature (dashed) (K), and (d) equivalent potential temperature (K). The abscissa is the distance from the cyclone center in km, and the ordinate is the height in km.

From the hydrostatic equation and the equation of motion, the difference between sea-level pressures at the center of the cyclone and at its edge can be determined: $\Delta p \approx p_0 g H \Delta T / RT^2$, where $p_0$ is the sea-level pressure and H is the height of the cyclone. Setting $p_0$ equal to 1000 hPa and taking H = 9500 m, $\Delta T$ = 3 K, T = 265 K, where $\Delta T$ is the height-averaged anomaly of temperature at the cyclone center and T is the height-averaged air temperature on the periphery, which are known from the simulation, we obtain $\Delta p \approx$ 14 hPa, an estimate close to the numerical result of 16 hPa.

At the isobaric surface, the radial temperature gradient is adjusted to the vertical shear of the azimuthal wind velocity through the thermal-wind approximation. The presence of a warm core in the eye of the cyclone means a decreased velocity in the eyewall with height, which corresponds to Fig. 6a. Apart from the warm core in the eye, there is a thermal inversion at 3 km, which separates dry air above and moist air below. This inversion suppresses deep convection in the eye, allowing only clouds in the boundary layer.

The azimuthally averaged distribution of equivalent potential temperature $\theta e(r, z)$ at 12:00 September 27 is shown in Fig. 7d. In the boundary layer, the temperature increases inward from the edge to the center. The maximum of $\theta e$ is at the center of the cyclone, which is explained by moist convection. We estimate the balance of θe for an air parcel that travels at a height of 500 m above sea level from the periphery of the cyclone toward the center. The temperature θe increases owing to the moisture flux and the sensible heat flux from the surface of the sea and decreases owing to the pressure drop. It can be shown that

$$\frac{\Delta \theta_e}{\theta_e} \approx \frac{L}{C_p \cdot T} \cdot \Delta q + \frac{\Delta T}{T} - \frac{R_a}{C_p}\frac{\Delta p}{p},$$

where L is the latent heat of evaporation, $C_p$ is the heat capacity of air at constant pressure, and $R_a$ is the gas constant for air. Replacing T, q, p, and θe by their mean values on the path of movement of the air parcel, we estimate the increment for equivalent potential temperature θe. From the simulation results at 500 m above sea level, θe = 324 K, T = 289.7 K, p = 957.3 hPa, and q = 10.8 g/kg at R = 150 km, and θe = 333.9 K, T = 291 K, p = 945.3 hPa, and q = 13.3 g/kg at R = 0. We find that the increment Δθ at the center of the eye is composed of 1.2 K from isothermal expansion, 1.5 K from sensible heat, and 7 K from latent heat, which add up to 9.7 K. This value is close to Δθ of 9.9 K obtained from the simulation. It is evident that the moisture flux from the sea surface plays an important role in increasing the equivalent potential temperature in the boundary layer. Then, this moisture is transferred upward in the eyewall by convection, with the result that the equivalent potential temperature increases in the eyewall (Fig. 7d). Thus, the surface fluxes of moisture and sensible heat are the energy source for convection in the eyewall.

At middle levels in the atmosphere, the equivalent potential temperature has minimum values in the eyewall and on the periphery of the cyclone because of low humidity. Above these levels, there is a small tongue of air with high θ, which is produced by subsidence of stratospheric air inside the eye.

We now discuss the mechanisms of the development and maintenance of secondary circulation. As was noted above, the first mechanism is the release of latent heat of condensation in the eyewall in deep convection, the resulting heating of air, and the enhancement of buoyancy. The higher buoyancy relative to the periphery of the cyclone intensifies the rise of air in the eyewall and, consequently, the low-level convergence. Another mechanism includes the surface friction of the azimuthal velocity component, a primary circulation. The rise of air produces the cyclonic vorticity of primary circulation, leads to the Ekman convergence owing to friction in the boundary layer, and gives rise to an additional vertical velocity at the top of the boundary layer.

Both theories of tropical cyclogenesis include both of these mechanisms, CISK and WISHE, which are discussed in detail in [24]. It is assumed that the larger contribution comes from the WISHE mechanism, in which an additional positive feedback between the azimuthal wind velocity and sensible and latent fluxes from the ocean surface plays a key role: increasing wind velocity leads to an increase in surface fluxes of heat, which spreads upward owing to convection in the eyewall, amplifies the air buoyancy, enhances the low-level convergence, and intensifies the primary circulation owing to the conservation of angular momentum. The results of the sensitivity experiments to be described below indicate that this mechanism is also crucial to the Black Sea midlatitude cyclone.

### 5. Momentum balance

One of the early assumptions in the theory of tropical cyclones was the hypothesis of hydrostatic and gradient balance [1, 27]. It is assumed that the central pressure of the cyclone is lower because of the warm core due to hydrostatics. The radial pressure gradient force is compensated by the centrifugal force and by the Coriolis force associated with the azimuthal circulation; i.e., the gradient balance is fulfilled. Thus, the state of the cyclone is determined by the temperature distribution, from which the distributions of pressure and of the azimuthal velocity can be obtained. The consequence of balance is a large lifetime of tropical cyclones, which far exceeds a typical time scale, the time of revolution. The deviations from these balances are assumed to be small. One of the unbalanced processes is the secondary circulation. It determines a slow change in the balanced state. During evolution, the tropical cyclone slowly changes from one balanced state to another. Other processes that destabilize the hydrostatic and gradient balance are fast. They include convection with large vertical accelerations and inertia – gravity waves (with consideration for strong radial and vertical shear of the azimuthal circulation).

The assumption of the hydrostatic and gradient balances has been conclusively confirmed in both field measurements [2] and numerical models [21, 29, 30].

The Black Sea cyclone examined in this paper resembles a tropical cyclone in terms of its structure and formation mechanism. Like the tropical cyclone, it is a long-lived phenomenon. Five days is a long period for typical midlatitude processes of a similar spatial scale. It is seen from the observations and from the numerical simulation that its evolution, growth, and decay occurred slowly. It can be supposed that the Black Sea cyclone, like a tropical cyclone, was well balanced. Therefore, it is interesting to estimate the accuracy of the gradient and hydrostatic balance for this cyclone and to estimate relative contributions of all forces to the equation of motion for the three momentum components: radial, azimuthal, and vertical ones. This makes it possible to determine mechanisms producing the primary azimuthal and secondary toroidal circulations and to estimate typical growth or decay rates for them.

*Balance of Radial Momentum*

We calculate the balance of radial momentum for the axisymmetric cyclone studied in this paper. In cylindrical pressure coordinates, the equation of motion for the radial velocity is written as

$$\frac{dV_r}{dt} = -g\frac{\partial h}{\partial r} + \frac{V_\theta^2}{r} + fV_\theta + F_r, \quad (1)$$

where $\frac{d}{dt} \equiv \frac{\partial}{\partial t} + V_r\frac{\partial}{\partial r} + \frac{V_\theta}{r}\frac{\partial}{\partial \theta} + V_z\frac{\partial}{\partial p}$ is the total derivative, g is the acceleration of gravity, h is the geopotential, $V_r$ and $V_\theta$ are the respective radial and azimuthal velocities in cylindrical coordinates, and $V_z$ is the vertical velocity in isobaric coordinates.

Equation (1) shows that the radial acceleration of an air parcel is determined by the following forces: the radial pressure gradient force $-g\frac{\partial h}{\partial r}$, the centrifugal force $\frac{V_\theta^2}{r}$, the radial Coriolis force $fV_\theta$, and the surface radial frictional force $F_r$. The Coriolis force component with the vertical velocity is omitted because the vertical velocity is two orders of magnitude less than the azimuthal velocity (Fig. 6).

Each variable a is represented as the sum of the azimuthal mean $\bar{a}$ and the wavelike (vortex) component $a'$, where the overbar means the azimuthal averaging and the prime denotes the deviation from the azimuthal mean. Equation (1) is averaged azimuthally, in which case the nonlinear terms yield the mean and vortex contributions $\overline{ab} = \bar{a}\bar{b} + \overline{a'b'}$. The vortex contributions of all of the nonlinear terms are small, particularly for azimuthal advection $\overline{\frac{V_\theta}{r}\frac{\partial V_r}{\partial \theta}} = \overline{\frac{V_\theta'}{r}\frac{\partial V_r'}{\partial \theta}} \approx 0$. Finally, the equation for the balance of radial momentum becomes

$$\frac{\partial \overline{V_r}}{\partial t} + \overline{V_r\frac{\partial V_r}{\partial r}} + \overline{V_z\frac{\partial V_r}{\partial p}} = -g\frac{\partial \overline{h}}{\partial r} + \frac{\overline{V_\theta^2}}{r} + \overline{fV_\theta} + \overline{F_r}. \quad (2)$$

Here, $\overline{V_r\frac{\partial V_r}{\partial r}}$ is the radial advection and $\overline{V_z\frac{\partial V_r}{\partial p}}$ is the vertical advection.

Figure 8 shows the terms of the balance equation for the radial momentum component at 12:00 September 27. The radial pressure gradient force, directed toward the cyclone center everywhere, has the largest magnitude (Fig. 8a). It reaches its maximum near the surface and decreases upward; the distribution along the radius has a maximum of –0.015 m/s$^2$ in the eyewall. The centrifugal force is directed outward from the center, is somewhat smaller than the pressure force, and has about the same distribution (not shown). The distribution of the Coriolis force is the same as that of the azimuthal velocity in Fig. 6c, but its magnitude in the eyewall is approximately one-fifth of the centrifugal force.

The sum of these three forces is illustrated in Fig. 8b. It is seen that the gradient balance is fulfilled with a sufficiently high accuracy, and the maximum value of this sum is 0.002 m/s$^2$, an order of magnitude less than the pressure gradient. This means that the Black Sea cyclone is a well-balanced system. The deviation from the gradient balance is positive almost everywhere, has a maximum in the eyewall, and decreases with height. This implies that a Lagrangian air parcel experiences a positive radial acceleration when it is lifted in the eyewall, which leads to the inclination of the eyewall outward with height, clearly seen in Fig. 6.

To estimate a local radial acceleration, we need to determine the radial and vertical advection of the radial velocity. Figure 8c shows the distribution of radial advection; it is much smaller than the preceding terms, and a maximum value is 0.0005 m/s$^2$. In the lower part of the inflow layer, radial advection is negative on the periphery of the cyclone, where the inflow accelerates, and reaches large positive values in the eyewall, where the inflow retards abruptly (Fig. 6c). In the upper layers of the outflow region, radial advection is positive at the inner edge of the eyewall due to the increased outflow rate and negative at the outer edge of the eyewall, where the radial velocity decreases with distance from the center.

The distribution of the vertical advection of the radial velocity is shown in Fig. 8d. The vertical advection has the same order of magnitude as the radial advection. It reaches large positive values in the eyewall region with a maximum of 0.0011 m/s$^2$ in the eyewall at the top of the boundary layer, where the radial velocity changes the sign with height and inflow is replaced by outflow. The vertical advection is positive in the central part of the eyewall, negative at the inner edge of the eyewall due to a decrease in the outflow rate with height at a fixed radius, and negative at the outer edge of the eyewall due to a change of sign of the radial velocity (Fig. 7c).

The sum of the three forces and of the two advection terms (with a minus sign) determines a local radial acceleration. The local radiation is shown in Fig. 8e. It is seen that advection has compensated more than half of the Lagrangian acceleration, and the residual term has a maximum value of 0.001 m/s$^2$, positive in the eyewall and decreasing with height. This means a local intensification of the radial outflow. In the boundary layer, the local radial acceleration takes on large negative values, which are associated with surface drag. Figure 8f shows the radial distribution of the radial component of wind stress, which is negative with a maximum value of 0.32 N/m$^2$ in the eyewall. If we take the boundary-layer depth to be 500 m, the acceleration will then be equal to – 0.001 m/s$^2$, a value close to that in Fig. 8d.

The main results of the analysis performed are the sufficiently exact fulfillment of the gradient balance and a positive Lagrangian acceleration in the eyewall, which leads to its inclination outward with height. Nonetheless, the maximum value of the Eulerian local acceleration of about 0.001 m/s$^2$ or 3 m/s per hour is yet sufficiently large to characterize long-term changes in the secondary circulation. For a more correct estimation of long-term tendencies in the radial velocity, it is probably necessary to average the balance equation over time and to take into account the vertical and horizontal friction in more detail.

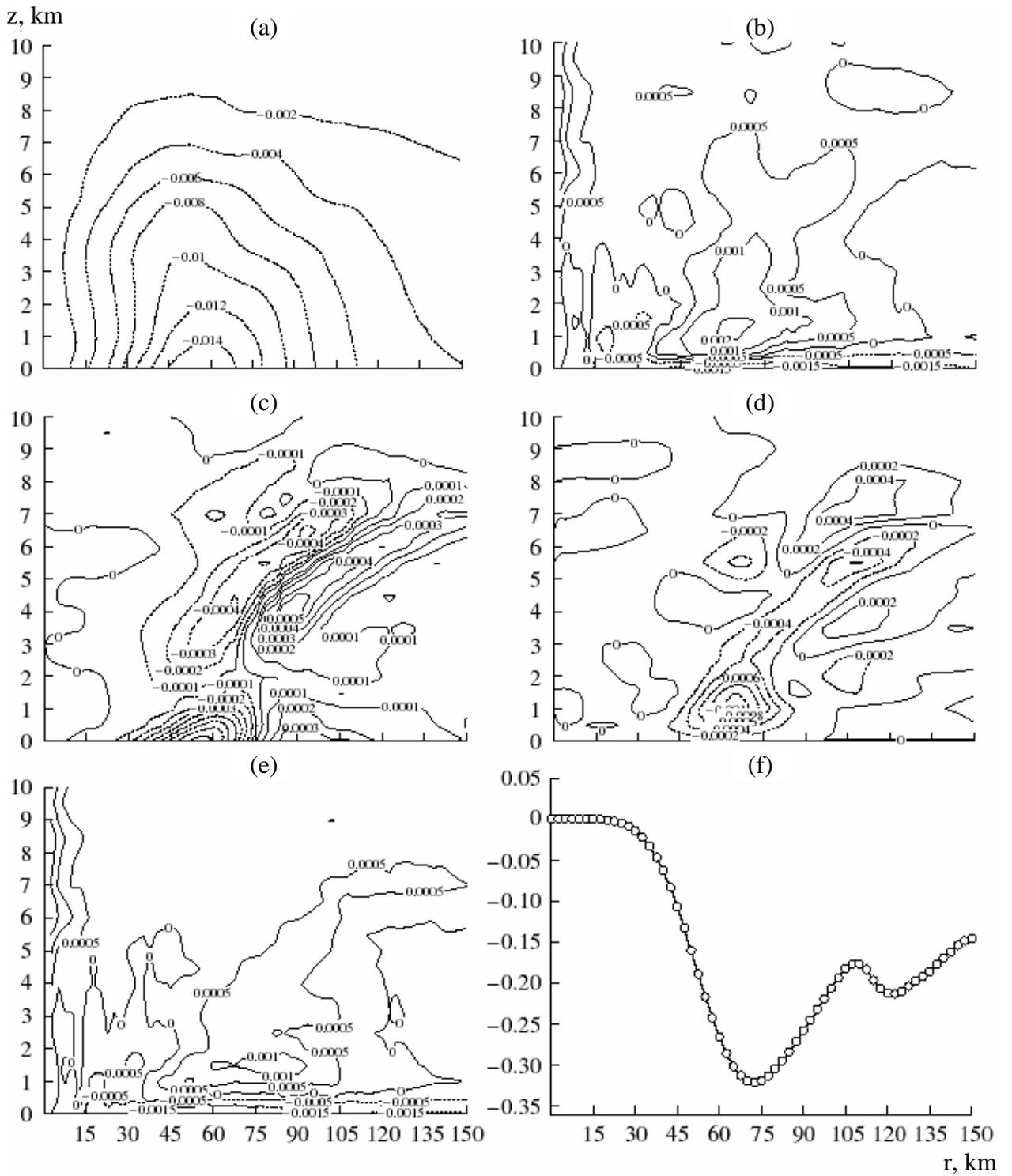

Fig. 8. Balance of the radial wind velocity at 12:00 September 27: (a) the radial pressure gradient force $-\overline{g\frac{\partial h}{\partial r}}$; (b) gradient balance $-\overline{g\frac{\partial h}{\partial r}} + \overline{\frac{V_\theta^2}{r}} + \overline{fV_\theta}$; (c) radial advection $\overline{V_r \frac{\partial V_r}{\partial r}}$; (d) vertical advection $\overline{V_z \rho g \frac{\partial V_r}{\partial p}}$; (e) $-\overline{g\frac{\partial h}{\partial r}} + \overline{\frac{V_\theta^2}{r}} + \overline{fV_\theta} - \overline{V_r \frac{\partial V_r}{\partial r}} - \overline{V_z \frac{\partial V_r}{\partial p}}$, values in (a) – (e) are in m/s²; and (f) radial surface wind stress (N/m²). Solid (dashed) lines are positive (negative) values.
Key: 1. km

*Balance of Azimuthal Momentum*

We now calculate the balance of the azimuthal velocity for an axisymmetric cyclone. The equation of motion for the azimuthal momentum is written as

$$\frac{dV_\theta}{dt} = -\frac{g}{r}\frac{\partial h}{\partial \theta} - \frac{V_r V_\theta}{r} - fV_r + F_\theta. \quad (3)$$

Equation (3) means that the azimuthal acceleration of an air parcel is determined by the following forces: the azimuthal pressure gradient force $-g\frac{\partial h}{\partial \theta}$, the inertial force $-\frac{V_r V_\theta}{r}$, the azimuthal Coriolis force $-fV_r$, and the boundary-layer azimuthal frictional force $F_\theta$. The Coriolis force component with the vertical velocity is omitted. Equation (3) is averaged azimuthally, and it is taken that $-\overline{\frac{g}{r}\frac{\partial h}{\partial \theta}} = -\frac{g}{r}\frac{\partial \overline{h}}{\partial \theta} = 0$  $\overline{\frac{V_\theta}{r}\frac{\partial V_\theta}{\partial \theta}} = \overline{\frac{V_\theta'}{r}\frac{\partial V_\theta'}{\partial \theta}} \approx 0$. Finally, the balance equation for the azimuthal momentum becomes

$$\frac{\partial \overline{V_\theta}}{\partial t} + \overline{V_r \frac{\partial V_\theta}{\partial r}} + \overline{V_z \frac{\partial V_\theta}{\partial p}} = -\frac{\overline{V_r V_\theta}}{r} - \overline{fV_r} + \overline{F_\theta}. \quad (4)$$

The different terms in the balance equation for the azimuthal momentum are presented in Fig. 9. The inertial force $-\frac{\overline{V_r V_\theta}}{r}$ and the Coriolis force on the right-hand side of (4) have a qualitatively similar distribution, with the Coriolis force in the eyewall being approximately one-fifth of the inertial force. For this reason, Fig. 9a shows their sum alone. This sum is positive in the lower layer of inflow and negative in the upper layers of outflow. Maximum positive values in the lower layer reach 0.0024 m/s$^2$, and negative values aloft are –0.0008 m/s$^2$. The sum of the inertial force and of the Coriolis force is directed rightward of wind velocity; therefore, it produces the cyclonic acceleration in the inflow layer and the anticyclonic acceleration in the outflow layer. This is one of the main mechanisms for the spin-up of the azimuthal circulation in the tropical cyclone due to the action of secondary circulation. The same has been demonstrated earlier, when the conservation of the absolute angular momentum was considered: the low-level convergence produces an increase in azimuthal velocity, while the upper-level divergence results in its decrease.

The distribution of the radial advection of azimuthal velocity is shown in Fig. 9b. It has large positive values of up to 0.0012 m/s$^2$ in the inflow region from the periphery to $R_{max}$, the radius of maximum wind, where the quantities $V_r$ and $\partial V_\theta/\partial r$ are both negative and large. At the inner edge of the eyewall, $\partial V_\theta/\partial r$ is reversed in sign, as is the radial advection. This means that radial advection leads to a local deceleration of rotation at the outer edge of the eyewall and to the acceleration of rotation at the inner edge, i.e., to the radial contraction of the cyclone. In the upper layers of the outflow region, radial advection has small negative values of –0.0002 m/s$^2$, because $V_r$ is positive and $\partial V_\theta/\partial r$ is negative, i.e., leads to an additional spin-up.

The distribution of the vertical advection of azimuthal velocity is shown in Fig. 9c. As might be expected, the vertical advection reaches large negative values of –0.0012 m/s$^2$ in the eyewall, because the vertical velocity is positive and the azimuthal velocity decreases with height. This is another important mechanism for the generation of the upper-level azimuthal circulation, i.e., the vertical advective transfer of azimuthal momentum.

The sum of these two forces and of the two advection terms (with the minus sign) is presented in Fig. 9d. It is a local azimuthal acceleration that is positive everywhere, which suggests the increase in tangential velocity, i.e., the spin-up of the cyclone. As can be seen from this figure, the line of the maximum local acceleration passes to the left of the line of maximum velocity. This indicates a tendency for a decrease in the cyclone radius. These tendencies may have induced small changes in the cyclone velocity and radius, which are discernible in Fig. 5.

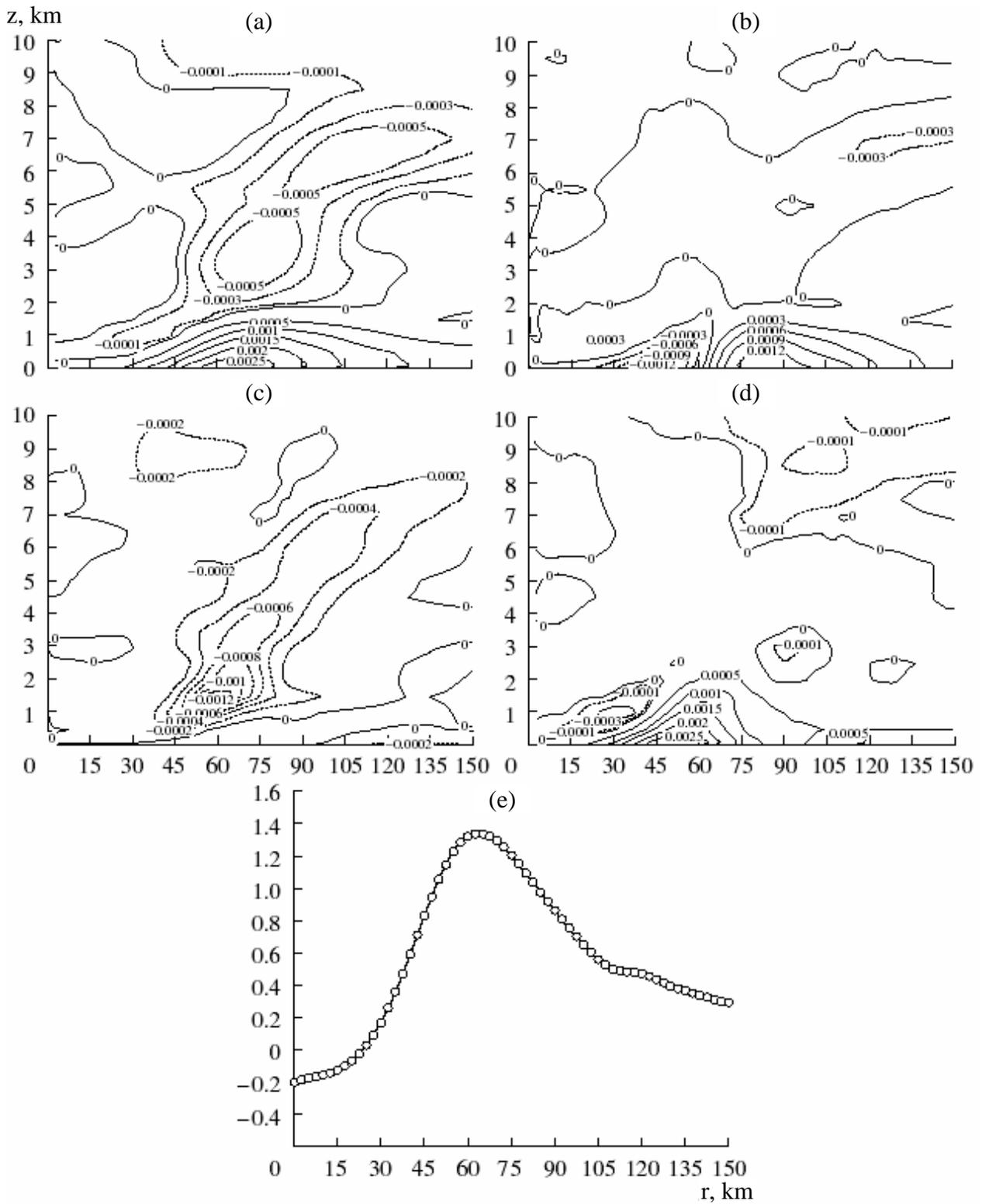

Fig. 9. Balance of the azimuthal wind velocity at 12:00 September 27: (a) the sum of the Coriolis force and the inertial force $-\dfrac{\overline{V_r V_\theta}}{r} - \overline{fV_r}$; (b) radial advection $\overline{V_r \dfrac{\partial V_\theta}{\partial r}}$; (c) vertical advection $\overline{V_z \dfrac{\partial V_\theta}{\partial z}}$, (d) $-\dfrac{\overline{V_r V_\theta}}{r} - \overline{fV_r} - \overline{V_r \dfrac{\partial V_\theta}{\partial r}} - \overline{V_z \dfrac{\partial V_\theta}{\partial p}}$, values in (a) – (d) are in m/s²; and (e) azimuthal component of wind stress in the boundary layer (N/m²). Solid (dashed) lines are positive (negative) values.
Key: 1. km

In the boundary layer, the eddy viscosity force, which retards the azimuthal circulation, has to be taken into account. Figure 9d shows the radial distribution of the azimuthal component of wind stress at the sea surface. It is qualitatively similar to the distribution of the local azimuthal acceleration in the boundary layer, with a maximum of 1.5 N/m$^2$ inside the radius of maximum winds. If we assume that this stress is distributed in a layer 500 m deep, the frictionally induced acceleration will be $-0.0024$ m/s$^2$, a value close to that in Fig. 9d.

*Balance of the Vertical Momentum Component*

We now calculate the balance of the vertical wind velocity for an axisymmetric cyclone. The simplified equation of motion for the vertical momentum in cylindrical pressure coordinates is written as

$$\frac{dV_z}{dt} = -g^2 \rho \frac{\partial h'}{\partial p} - g \frac{\rho'}{\rho} - g(q_c + q_r + q_i + q_s + q_g) + F_z, \quad (5)$$

where $\rho$ is the air density; $\rho'$ and $h'$ are perturbations of density and geopotential, respectively, relative to the horizontally averaged values; and $q_c$, $q_r$, $q_i$, $q_s$, and $q_g$ are the mixing ratios of cloud droplets, raindrops, cloud ice, snow, and graupel, respectively. Equation (5) means that the vertical acceleration of an air parcel is determined by the following forces: the force of the vertical gradient of pressure perturbations $-g^2 \rho \frac{\partial h'}{\partial p}$, the buoyancy force $-g \frac{\rho'}{\rho}$, the condensed-water weight $g(q_c+q_r+q_i+q_s+q_g)$, and the boundary-layer vertical frictional force $F_z$. The vertical component of the Coriolis force is omitted. By averaging Eq. (5) azimuthally, we finally obtain

$$\frac{\partial \overline{V_z}}{\partial t} + \overline{V_r \frac{\partial V_z}{\partial r}} + \overline{\frac{V_\theta}{r} \frac{\partial V_z}{\partial \theta}} + \overline{V_z \frac{\partial V_z}{\partial p}} = -\overline{g^2 \rho \frac{\partial h'}{\partial p}} + \overline{g \frac{\rho'}{\rho}} - \overline{g(q_c + q_r + q_i + q_v)} + \overline{F_z}. \quad (6)$$

The main forces in Eq. (6) are the buoyancy force and the vertical pressure gradient force, which compensate each other. The buoyancy force is positive in the cyclone if r < 100 km and reaches a maximum of 500 m s$^{-1}$ h$^{-1}$ in the eye of the cyclone at 550 hPa, where the warm core is most heated. The residual terms in the eyewall are one or two orders of magnitude smaller than the buoyancy force. It means that the hydrostatic approximation in the eyewall is applicable with a high degree of accuracy. However, in the eyewall, where the vertical acceleration is on the same order of magnitude as the buoyancy force, the hydrostatic approximation in inapplicable. The weight of the condensed water has no large effect on the hydrostatic equilibrium because it is one or two orders of magnitude smaller than the buoyancy force. The advective terms and the local derivative on the left-hand side of (6) are negligible because they are on the order of 10$^{-5}$, i.e., three or four orders of magnitude smaller than the buoyancy force.

The main finding of the analysis of the balance of vertical momentum is that the hydrostatic approximation is fulfilled with high accuracy and the residual terms are small.

### 6. Sensitivity experiments

In order to identify mechanisms responsible for the formation of the Black Sea cyclone, numerical experiments have been performed, and their results are shown in Fig. 10. In each of the four experiments, the simulation outputs were compared with the results of the control run described above.

*Removal of Latent Heat Release*

It is known that the most important process in tropical cyclones for the growth of a vortex is the release of latent heat of condensation during convection in the eyewall. To confirm the importance of this mechanism for the Black Sea cyclone, the release of latent heat of condensation was removed from the numerical experiment. As can be seen in Fig. 10a, the cyclone failed to be reproduced. Thus, the release of latent heat, as might be expected, is one of the key processes in the formation of the cyclone.

*Removal of Surface Heat Fluxes*

All studies of tropical cyclones with the help of numerical models show the importance of the latent (and possibly sensible) heat flux at the sea surface, as was suggested in [31]. In [26], it was assumed that the intensification and development of tropical cyclones occur solely owing to the heat fluxes from the sea surface that are induced by the cyclones themselves and that even the convective available potential energy of an undisturbed state gives no significant contribution. Later in [32], with the help of a simplified model, it was shown that a tropical cyclone may develop in the atmosphere neutrally stable with respect to cumulus convection. As the cyclone of interest resembles a tropical cyclone, we estimate how the fluxes of sensible and latent heat from the sea surface influence its development.

The numerical experiment without these fluxes produced only a very weak cold-core cyclone that filled by September 28. The depth of the cyclone with no heat fluxes decreased substantially relative to the depth of the control cyclone, with a depth difference of 15 hPa (Fig. 10b). The surface wind speed also decreased by a factor of 2.5. The fact that the cyclone originated indicates the presence of the background convective available potential energy in the initial conditions. Moreover, the temperature at the center of the cyclone decreased significantly in this experiment. In contrast to the control cyclone, the temperature of the core in the cyclone with no heat fluxes was $-1$ to $-2.5°$ lower than the temperature of its environment.

Thus, the surface heat fluxes appear to play an important role in storm development. This result agrees with the WISHE theory [26], in which the surface heat fluxes are assumed to be the main mechanism of positive feedback in the air – sea interaction processes. In connection with this, the following experiment was performed to estimate the sensitivity of the vortex to the variation in positive feedbacks in the ocean – atmosphere system. The simplest way of estimating such positive feedbacks was to impose a restriction on the surface wind speed.

*Restriction on the Surface Wind Speed*

We limit a positive feedback between the heat fluxes and the wind speed by restricting the wind speed in the simulation of heat fluxes to 10 m/s, i.e., by reducing the nonlinear interaction between the heat fluxes and the circulation of the vortex. This procedure does not eliminate the interaction completely; the heat fluxes also increase owing to the increased air – sea temperature difference and to the increased difference in specific humidity at the surface and at the top of the boundary layer.

The following deviations from the control run have been obtained in the numerical experiment. First, the depth of the cyclone decreased: the difference between minimum central pressures of the control cyclone and of the cyclone with the limited heat fluxes reaches 7 hPa. Second, the track of the cyclone has changed significantly: the cyclone does not wander any longer in the limited southwestern part of the Black Sea, but, traveling along the south coast, moves eastward (Fig. 10c). Therefore, restricting the surface wind speed has produced a decrease in the heat fluxes. The heat fluxes decreased most significantly in the cyclone and especially under the eyewall. In the ambient atmosphere, where the surface wind speed was rarely above 10 m/s, no changes have occurred.

*Removal of Cooling due to Evaporation of Hydrometeors*

In this experiment, the cooling due to the evaporation of cloud droplets and raindrops has been removed. As a result, the cyclone was very vast and deep because of a strong unrealistic overheating (Fig. 10d). It was also located in a different area of the sea. This experiment points to the need to correctly parametrize physical processes in the simulation of the cyclones in which convection is an important factor.

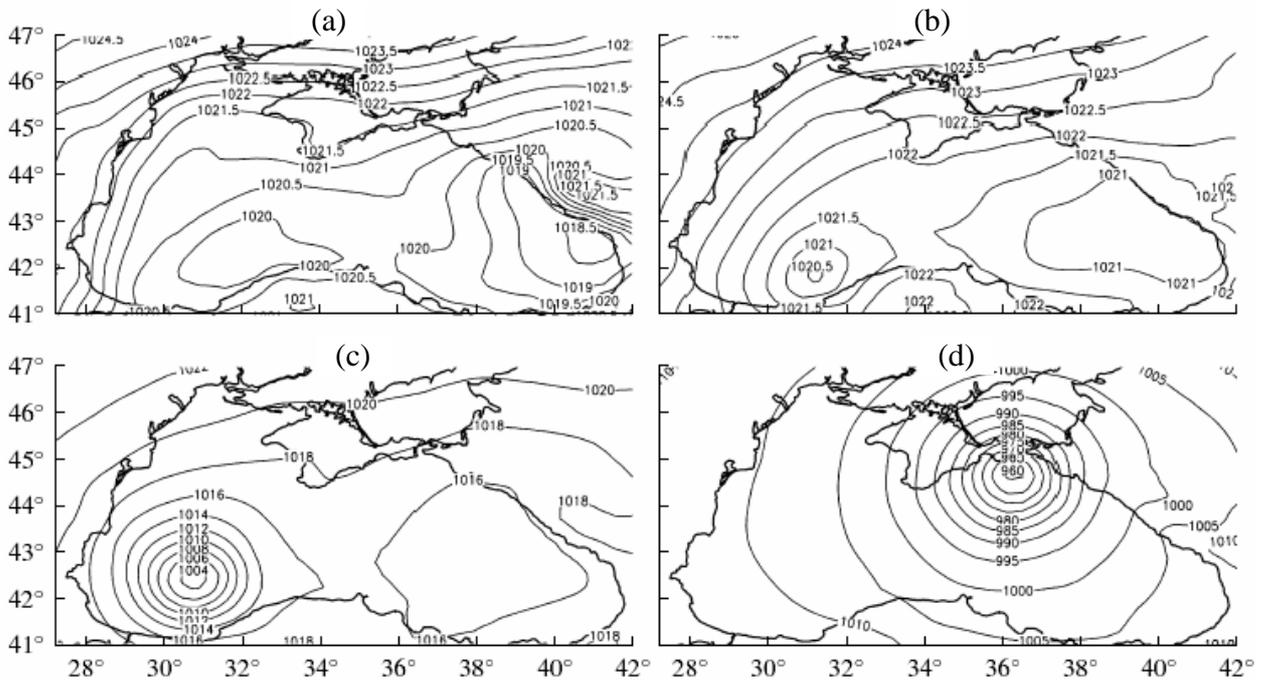

Fig. 10. Results obtained from sensitivity experiments. Sea-level pressure (hPa) at 12:00 September 27 with removal of (a) latent heat release, (b) surface heat fluxes, (c) restriction of surface wind speed, and (d) with no cooling due to evaporation of hydrometeors.

**Conclusions**

The paper presents results of the simulation of a quasi-tropical cyclone, rarely observed in the Black Sea region, which developed in early September 2005. The MM5 nonhydrostatic regional atmospheric circulation model was used, with the initial and boundary conditions taken from the operational global analysis.

The model has successfully reproduced the main properties of the cyclone and provided a detailed analysis of its structure, evolution, and physical properties.

First, the model has reproduced the evolution of the cyclone from September 25 to September 29, i.e., its growth, mature stage, and decay. Second, the model has described the track of the cyclone center: for five days, the vortex was wandering slightly in the southwestern part of the Black Sea. Third, the near-axisymmetric shape of the vortex with spiral cloud bands has been simulated. Fourth, the sizes of the cyclone were obtained to be close to the observed ones: the radius of maximum winds of 60 km and the radius of the cloud system equal to 150 km. Finally, the model has simulated the maximum surface wind speed in excess of 20 m/s.

In addition, the model has reproduced other properties of the cyclone that defy measurement. In particular, these properties include the primary circulation, with a surface maximum of the azimuthal wind velocity and with a gradual decay upward to the 300-hPa level, the secondary circulation with the low-level convergence, the rise of air in the eyewall, and upper-level divergence. Numerical simulations have shown that the cyclone has a cloud-free eye, where the air subsides, and a warm core at the center with an overheating of 3°C. Clouds and precipitation are concentrated in the eyewall consisting of several spiral bands. The total flux of sensible and latent heat reached very large values, up to 1000 W/m$^2$. Maximum values of the equivalent potential temperature were about 350 K.

The components of the balance of the radial, azimuthal, and vertical momentum in the cyclone have been estimated, which confirmed that the hydrostatic and gradient balances were fulfilled in the mature stage of the quasi-tropical cyclone. Numerical sensitivity experiments have been performed to demonstrate that the two important physical processes generating tropical cyclones, i.e., the transfer of sensible and latent heat from the ocean to the atmosphere and the transfer and release of moisture and heat in the upper atmosphere, are also crucial to the development of the Black Sea cyclone. Thus, it can be regarded with confidence as an extensively investigated tropical cyclone, although not all necessary requirements for the formation of tropical cyclones were satisfied. In particular, the commonly accepted requirement for the tropical cyclones to develop is the water temperature above 26 – 27°C, whereas the Black Sea cyclone originated at a much lower water temperature of 23°C.

**References**

1. K. A. Emanuel, "Tropical Cyclones", Annu. Rev. Earth Planet. Sci. V. 31, P. 75 – 104 (2003).
2. R. A. Anthes, "Tropical Cyclones: Their Evolution, Structure and Effects", Meteorological Monographs No. 41 (American Meteorological Society, Boston, 1982).
3. V. Homar, R. Romero, D. J. Stensrud, et al., "Numerical Diagnosis of a Small, Quasi-Tropical Cyclone over the Western Mediterranean: Dynamical vs. Boundary Factors", Q. J. R. Meteorol. Soc. V. 129, P. 1469 – 1490 (2003).
4. O. Reale and R. Atlas, "Tropical Cyclone-Like Vortices in the Extratropics: Observational Evidence and Synoptic Analysis", Weather Forecast. V. 16, P. 7 – 34 (2001).
5. K. Lagouvardos, V. Kotroni, S. Nickovic, et al., "Observations and Model Simulations of a Winter Sub-Synoptic Vortex over the Central Mediterranean", Meteorol. Appl. V. 6, P. 371 – 383 (1999).
6. L. Pytharoulis, G. C. Craig, and S. P. Ballard, "Study of the Hurricane-Like Mediterranean Cyclone of January 1995", Phys. Chem. Earth (B) V. 24, P. 627 – 632 (1999).
7. L. Pytharoulis, G. C. Craig, and S. P. Ballard, "The Hurricane-Like Mediterranean Cyclone of January 1995", Meteorol. Appl. V. 7, P. 261 – 279 (2000).
8. K. A. Emanuel, "Genesis and Maintenance of "Mediterranean Hurricanes"", Adv. Geophys. V. 2, P. 217 – 220 (2005).
9. J. A. Ernst and M. Matson, "A Mediterranean Tropical Storm? ", Weather V. 38, P. 332 – 337 (1983).
10. R. J. Reed, Y.-H. Kuo, M. D. Albright, et al., "Analysis and Modeling of a Tropical-Like Cyclone in the Mediterranean Sea", Meteorol. Atmos. Phys. V. 76, P. 183 – 202 (2001).
11. F. Meneguzzo, M. Pasqui, G. Messeri, and M. Rossi, "High-Resolution Simulation of a Deep Mediterranean Cyclone Using RAMS Model", Unpublished Manuscript (2001).
12. E. Rasmussen, "A Case Study of a Polar Low Development Over the Barents Sea", Tellus V. 37A, P. 407 – 418 (1985).
13. E. Rasmussen and J. Turner, "Polar Lows. Mesoscale Weather Systems in the Polar Regions", (Cambridge Univ. Press, Cambridge, 2003).
14. K. A. Emanuel and R. Rotunno, "Polar Lows As Arctic Hurricanes", Tellus A V. 41, P. 1 – 17 (1989).
15. http://poet.jpl.nasa.gov/
16. V. V. Efimov, S. V. Stanichnyi, M. V. Shokurov, and D. A. Yarovaya, "Observation of a Quasi-Tropical Cyclone over the Black Sea", Meteorol. Gidrol. (2007) (in press).
17. http://dss.ucar.edu/datasets/ds083.2/
18. K. A. Emanuel, "Atmospheric Convection", (Oxford Univ. Press, Oxford, 1994).
19. G. A. Grell, J. Dudhia, and D. R. Stauffer, "A Description of the Fifth-Generation Penn. State/NCAR Mesoscale Model (MM5) ", NCAR Technical Note (1995).
20. J. Dudhia, D. Gill, Guo Yong-Run, and K. Manning, "PSU/NCAR Mesoscale Modeling System. Tutorial Class Notes and User's Guide: MM5 Modeling System Version 3", NCAR Tutorial Notes (2005).

21. C. A. Davis and L. F. Bosart, "Numerical Simulations of the Genesis of Hurricane Diana (1984). Part I: Control Simulation", Mon. Weather Rev. V. 129, P. 1859 – 1881 (2005).

22. Liu Yubao, Zhang Da-Lin, and M. K. Yau, "A Multiscale Numerical Study of Hurricane Andrew. Part I: Explicit Simulation and Verification", Mon. Weather Rev. V. 125, P. 3073 – 3093 (1997).

23. J. G. Charney and A. Eliassen, "On the Growth of the Hurricane Depression", J. Atmos. Sci. V. 21, P. 68 – 75 (1964).

24. G. C. Craig and S. L. Gray, "CISK or WISHE As the Mechanism for Tropical Cyclone Intensification", J. Atmos. Sci. V. 53, P. 3528 – 3540 (1996).

25. E. A. Hendriks, M. T. Montgomery, and C. A. Davis, "Role of "Vortical" Hot Towers in the Formation of Tropical Cyclone Diana (1984) ", J. Atmos. Sci. V. 61, P. 1209 – 1232 (2004).

26. K. A. Emanuel, "An Air – Sea Interaction Theory for Tropical Cyclones. Part I. Steady State Maintenance", J. Atmos. Sci. V. 43, P. 585 – 604 (1986).

27. K. A. Emanuel, "The Theory of Hurricanes", Annu. Rev. Fluid Mech. V. 23, P. 179 – 196 (1991).

28. Liu. Yubao, Zhang. Da-Lin, and M. K. Yau, "A Multiscale Numerical Study of Hurricane Andrew (1992). Part II: Kinematics and Inner-Core Structures", Mon. Weather Rev. V. 127, P. 2597 – 2616 (1999).

29. Liu. Yubao, Zhang Da-Lin, and M. K. Yau, "A Multiscale Numerical Study of Hurricane Andrew (1992). Part III: Dynamically Induced Vertical Motion", Mon. Weather Rev. V. 128, P. 3772 – 3788 (2000).

30. Liu. Yubao, Zhang. Da-Lin, and M. K. Yau, "A Multiscale Numerical Study of Hurricane Andrew (1992). Part IV: Unbalanced Flows", Mon. Weather Rev. V.129, P. 92 – 107 (2001).

31. E. Kleinschmidt, "Gundlagen Einen Theorie Des Tropischen Zyklonen", Arch. Meteorol. Geophys. Bioklimatol. Ser. A V.4, P. 53 – 72 (1951).

32. R. Rotunno and K. A. Emanuel, "An Air – Sea Interaction for Tropical Cyclones. Part II: Evolutionary Study Using Nonhydrostatic Axisymmetric Numerical Model", J. Atmos. Sci. V. 44, P. 542 – 561 (1987).